\documentstyle[aps,prb,twocolumn]{revtex}
\input{epsf}
\newlength{\colwi}
\begin{document}
\twocolumn[\hsize\textwidth\columnwidth\hsize\csname@twocolumnfalse\endcsname
\draft
\preprint{\today}
\title{Thouless numbers for few-particle systems\\
with disorder and interactions}

\author{Dietmar Weinmann$^{1,2}$, Jean-Louis Pichard$^1$ and Yoseph Imry$^3$}
\address{1:~CEA, Service de Physique de l'Etat Condens\'e,
         Centre d'Etudes de Saclay,
         91191 Gif-sur-Yvette Cedex, France\\
         2:~Institut f\"ur Physik,
         Universit\"at Augsburg,
         Memminger Str.~6,
         86135 Augsburg, F.R.~Germany\\
         3:~Weizmann Institute of Science,
         Department of Condensed Matter Physics,
         76100 Rehovot, Israel}
\maketitle
\begin{abstract}
 Considering $N$ spinless Fermions in a random potential, we study how a short 
range pairwise interaction delocalizes the $N$-body states in the basis of the 
one-particle Slater determinants, and the spectral rigidity of the $N$-body 
spectrum. The maximum number $g_N$ of consecutive levels exhibiting the 
universal Wigner-Dyson rigidity (the Thouless number) is given as a function of
the strength $U$ of the interaction for the bulk of the spectrum. In the dilute
limit, one finds two thresholds: When $U<U_{c1}$, there is a perturbative 
mixing between a few Slater determinants (Rabi oscillations) and 
$g_N \propto |U|^{P} <1$, where $P=N/2$ (even $N$) or $(N+1)/2$ (odd $N$). When
$U=U_{c1}$, the matrix element of a Slater determinant to the ``first 
generation'' of determinants directly coupled to it by the interaction is 
of the order of the level spacing of the latter determinants, 
$g_N \approx 1$ and the level spacing distribution exhibits a crossover 
from Poisson to Wigner, related
to the crossover between weak perturbative mixing and effective golden-rule 
decay. Moreover, we show that the same $U_{c1}$ signifies also the breakdown of
the perturbation theory in $U$. For $U_{c1}<U<U_{c2}$, the states are extended 
over the energetically nearby Slater determinants with a non-ergodic 
hierarchical structure related to the sparse form of the  Hamiltonian. Above a 
second threshold $U_{c2}$, the sparsity becomes irrelevant, and the states are 
extended more or less ergodically over $g_N$ consecutive Slater determinants. A
self-consistent argument gives $g_N \propto U^{N/(N-1)}$. We compare our 
predictions to a numerical study of three spinless Fermions in a disordered 
cubic lattice. Implications for the interaction-induced $N$-particle 
delocalization in real space are discussed. The applicability of Fermi's golden
rule for decay in this dilute gas of "real" particles is compared with the one 
characterizing a finite-density Fermi gas. The latter is related to the 
recently suggested Anderson transition in Fock space.

\end{abstract}
\pacs{PACS numbers:
 72.15,
%Electronic conduction in metals and alloys; Collective modes
 73.20}]
%Electron states in low dimensional structures; Delocalization processes

\section{Introduction}\label{Intr}

For non-interacting electrons the Thouless energy $E_{\rm c}$ has proven to be 
a very relevant energy scale for several physical properties. The related 
``Thouless number'' $g_1 = E_{\rm c} / \Delta_1$, where $\Delta_1$ is the 
single-particle level spacing at the Fermi energy, plays an important role. For
the disordered case, $E_{\rm c} = \hbar D / L^2$, $D$ being the diffusion 
constant of the electrons and $L$ the relevant sample length. In this case, the
Thouless number $g_1$ is equal \cite{Thou} to the dimensionless conductance, 
i.e. the conductance in units of $e^2 / \hbar$. This important relationship is 
at the basis of the scaling theory of localization\cite{aalr}, which has been 
quite successful in describing transport in disordered metals for 
non-interacting electrons.

$E_{\rm c}$ is also an important energy scale for the spectral correlations of
diffusive non-interacting particles in a random potential. It was found by
Altshuler and Shklovskii\cite{AS} that the usual random-matrix correlations
\cite{Efet} of the density of states at different energies $E$ and $E'$ hold
only when the relative energy $|E - E'|$ is smaller than $E_{\rm c}$. This
means that one has only $g_1$ consecutive one-particle levels which exhibit the
universal Wigner-Dyson rigidity. For $|E - E'| \gtrsim E_{\rm c}$ a novel
spectral correlation function  was obtained which depends on the dimensionality
and the diffusion constant. This new dependence and the crossover associated
with it follow rather easily\cite{Arg} from a semiclassical theory for the
spectral correlations due to Berry\cite{Berry}.

When electron-electron interactions are introduced\cite{AA}, a single-electron
(or hole) excitation with an energy $\epsilon$ acquires a finite width
$\Gamma_{\rm sp}(\epsilon)$\footnote{The width of the single-quasi-particle 
excitation exists only when the energy of the excitation is high enough, namely
above a threshold $\epsilon^*$, and the golden-rule formulation for the decay 
is valid. An analysis of this question either by using \cite{Mor,Cur} the 
formulation of Ref.~\cite{BM}, or via a correspondence with localization on a 
Cayley tree\cite{agkl}, shows that this crossover energy $\epsilon^*$ is of the
order of $\sqrt{E_{\rm c} \Delta_1} \sim \Delta_1 \sqrt{g_1}$. Note that since
$\sqrt{E_{\rm c} \Delta_1} \ll E_{\rm c}$, the level width is well defined and
valid on the scale $E_{\rm c}$ and for a large window below it. In
Ref.~\cite{agkl} further extremely interesting results were obtained to which
we shall return later.}. This width obviously increases with $\epsilon$. It has
been calculated for an {\em isolated} metallic ($g_1 \gg 1$) compact quantum 
dot in Ref.~\cite{SIA}, where it turns out that at the Thouless energy 
$E_{\rm c}$ this width becomes comparable to $\Delta_1$ and the 
single-quasi-particle excitations can no longer be resolved. Thus the number of
single-particle levels that can be resolved is of the order of the Thouless
number $g_1$. This result which agrees with the experimental findings of
Ref.~\cite{Sivan} is universal and does not depend on material parameters, nor
on the dimensionality of the dot.

 The problem of interacting particles in a random potential is of great
fundamental interest. In particular, the suggestion \cite{shepelyansky} that
some states of two interacting particles (TIP) in a random potential may be
less strongly localized than each particle separately has recently caused much
interest\cite{imry,fmgpw,wmgpf,wp,fmgp,Moriond,jacquod,oppen,oppen2}. This idea
can be understood within the scaling theory based on the Thouless 
picture\cite{imry}. According to this, the delocalization of the particular TIP
states from one block of size $L_1$ (the one-particle localization length) to 
the neighboring block follows from their having an interaction dependent 
Thouless number $g_2 \sim E_{\rm c2}/ \Delta_2$ much larger than that of the 
single-particle states $g_1$. Here, $\Delta_2$ is the two-particle level 
separation at the given energy and the corresponding Thouless energy 
$E_{\rm c2}$ was first identified~\cite{imry} to the interaction dependent 
decay rate between neighboring blocks.  As in the single-particle case, the 
multiple role played by the Thouless energy, as discussed above, immediately 
suggested that this TIP Thouless parameter will also be relevant to the level 
correlation problem. This is based on the general qualitative picture. While 
the TIP-spectrum without interaction contains only hidden one-particle 
correlations appearing on energy scales larger than $\Delta_1$ and is close to 
uncorrelated levels on lower energy scales, the interaction re-establishes the 
universal Wigner-Dyson rigidity up to the energy $E_{c2} \equiv E_U$ which 
depends on the strength $U$ of the interaction. In the {\it localized} regime 
($L>L_1$), this was formally described by a nonlinear $\sigma$-model for the 
TIP problem, as presented in Ref.~\cite{fmgp}. The latter gives a theoretical 
foundation for the scaling picture for TIP on equal footing to that for 
non-interacting particles. In the {\it metallic} regime $(L<L_1)$, a study of 
the TIP-level statistics~\cite{wp} confirmed that $E_U$ gives also the 
characteristic energy scale up to which the TIP-spectrum exhibits the universal
Wigner-Dyson rigidity. This was qualitatively explained by mapping~\cite{wp} 
the TIP-Hamiltonian onto a Gaussian matrix model with preferential basis.

  We see that the scaling properties for interacting particles can thus be
studied via the spectral correlations of their levels {\em in the metallic
regime}. This is an extremely useful observation. The metallic regime is easier
to study both analytically, where reliable methods exist, and numerically. In
the latter case, the necessity to go to very large system sizes larger than the
localization length with weak disorder in low dimensions is eliminated thereby.
Since the study of two interacting particles is only the first step towards the
treatment of a more realistic many-body system, it is highly desirable to 
increase the number of particles. Even a modest program of going from two to 
three, four and larger numbers of interacting particles can be best 
accomplished by analyzing the Thouless parameters in the metallic regime for 
rather small system sizes. This is the strategy we adopt in this paper.

  It was mentioned before (footnote 1) that when a state is coupled to a
quasi-continuum, the golden rule expression for its width starts to be valid
only when the coupling is strong enough, or the density of the final states is
high enough. The crossover between perturbative mixing (Rabi regime) and
effective decay in fact occurs when the typical matrix element of the coupling
becomes larger than the mean level spacing of the accessible states
\cite{Mor,Cur,wp}. An equivalent condition is that the golden-rule width be
larger than the final level spacing. This very general crossover, which becomes
a phase transition in the appropriate ``thermodynamic limit'' is the essence of
delocalization in the usual Thouless scaling theory for a single particle. It 
applies to two-particle delocalization\cite{imry} and should likewise describe 
delocalization for $N$ particles. The Hilbert-space transition found in 
Ref.~\cite{agkl} is another example. In this case one gets a proper transition
by the hierarchical coupling to higher and higher numbers of quasi-particle 
excitations. In the work presented here, as in Ref.~\cite{wp}, this transition 
is observed numerically as a function of the interaction strength $U$. When $U$
is weak, it can couple only a few very close quasi-degenerate states and leads 
at most to Rabi-type oscillations between adjacent levels. When $U$ is larger 
than a certain threshold $U_{c1}$ (or at larger excitation energy), many 
non-interacting states are coupled and Fermi's golden rule describes the 
spreading width of a non-interacting state over the (quasi-continuum) of other 
non-interacting states which are nearby in energy. $U_{c1}$ is also the 
crossover interaction between Poisson and Wigner-Dyson statistics for the 
spectral fluctuations. We show that the same $U_{c1}$ signifies also the 
breakdown of the perturbation theory in $U$. Above a higher threshold $U_{c2}$,
the states are ergodically mixed and $g_N$ is suggested to increase like 
$U^{N/(N-1)}$. 

 Most of this paper will be concerned with the three-particle problem. Simple
analytical arguments will be presented for the behavior of the spectral
correlations in a small diffusive quantum dot, and compared to a numerical
study. Thus, we work in this paper in the metallic regime and do not {\it
directly} study the delocalization in real space for stronger disorder, when 
the one-particle states are localized. However, some remarks will be eventually
made on the implication of this picture to interaction-induced delocalization 
in real space and on its generalization to quasi-particle excitations in a 
degenerate metallic Fermi system. In particular, the basic delocalization 
mechanism  discussed in the original locator expansion of Anderson was the 
divergence of perturbation theory around the initial localized states. It will 
be shown that a seemingly analogous divergence can be identified in the 
perturbation theory in the interaction, around the noninteracting states. Here,
this divergence signifies (as is also true in the Anderson localization case) 
the onset of Wigner-Dyson correlations in the full spectrum, where the basis of
noninteracting eigenstates becomes well-mixed due to the interactions. A 
similar process appears in the recent work~\cite{agkl} of Altshuler et al., 
using an analogous expansion for the quasiparticle excitations in a degenerate 
Fermi system, decaying by emitting electron-hole pairs. In the three cases of 
the Anderson delocalization in real space and the delocalization processes 
found due to interactions in the Hilbert space of wavefunctions, the basic 
condition for the transition is very similar. It demands that the matrix 
element of a state to the "first generation" of states directly coupled to it 
by the interaction, be comparable to the level spacing of the latter states.

\section{$N$-body Hamiltonian in the Fock basis}

  In the presence of interactions, it is convenient to consider the $N$-body 
system in a certain Fock basis. Since we use this terminology in a slightly
unusual way, let us make precise what we mean by Fock basis. We consider the
one-particle states which take into account exactly the kinetic energy, the
random electrostatic potential seen by the electron, the chaotic or the
integrable dynamics yielded by the boundaries in a ballistic billiard etc., and
we use the exact one-particle states to build up the Slater determinants which
we refer to as the Fock states. Therefore, by Fock basis we just mean the
eigenbasis of the $N$-body Hilbert space in which the system Hamiltonian is
diagonal at $U=0$.

 Moreover, we do not focus on the low excitation energies, (i.e.~on the
restricted space available from the Fermi vacuum by successive applications of
quasi-particle creation operators) but rather to higher energies in the bulk of
the $N$-body spectrum. Therefore, in contrast to Ref.\cite{agkl}, the parameter
in our study is not the excitation energy of an extra quasi-particle above the
Fermi sea, but the strength $U$ of the interaction, at a given total energy 
chosen close to the band center of the $N$-body spectrum. Another important
difference between this study and the problem considered in Ref.\cite{agkl} is 
that we have in mind the {\it dilute limit} \footnote{ This dilute limit 
strictly means that $N/M \rightarrow 0$ when $M \rightarrow \infty$, where 
$M=L^d$ is the number of sites in a tight-binding model. This should be 
distinguished from the finite density limit where $N/M$ is a constant and the 
Fermi system is degenerate at zero temperature}, where the number of ``real'' 
particles is arbitrary, but nevertheless of zero density. Therefore, we have 
not in this study a Fermi vacuum from which an arbitrary large number of 
quasi-particles can be indefinitely created.

 In this Fock basis, the Hamiltonian with interaction is a random matrix with
preferential basis. For the sake of simplicity, we assume that the one-particle
states are more or less uniformly extended inside the sample (no one-particle
localization); i.e.~the Hamiltonian without interaction  ${\cal H}_0$ is a sum 
of one-particle Hamiltonians which can be described by a random matrix being 
statistically invariant under the orthogonal transformation $O(M)$. $M=L^d$ is 
the number of considered sites for a sample size $L$. ${\cal H}_0$, which 
contains the kinetic and potential energy of the particles is combined with a 
two-body interaction of the form
\begin{eqnarray} \label{hamiltonian2}
{\cal H}_{\rm int} = { 1 \over 2} \sum_{ijkl}
 \langle ij | \hat{U} | kl \rangle c^{+}_{i} c^{+}_{j} 
c^{\phantom{+}}_{l} c^{\phantom{+}}_{k} \, ,
\end{eqnarray}
where $|i\rangle $ denotes the Wannier function localized at the $i^{\rm th}$
site, $c^{\phantom{+}}_i/c^{+}_i$ destroys/creates a particle on site $i$, and
\begin{equation}
\hat{U} = \sum_{i<j} |ij\rangle u_{ij} \langle ij|
\end{equation}
is a local interaction of strength measured by a parameter $U$. We will give
estimates assuming on-site interaction in the following, but the interaction
should be obviously extended to nearest neighbors in the case of spinless
Fermions. We write the Hamiltonian in the basis of the
$n_{\rm tot}=M!/(N!(M-N)!)$ Slater determinants (Fock states), which are
antisymmetrized products $|A_n>=|\alpha_1,\ldots,\alpha_N>$
($n=1,\dots ,n_{\rm tot})$ of one-particle eigenfunctions $|\alpha_j >$.
${\cal H}_0$ is a diagonal matrix with the different possible sums
$\sum_{I=1}^{N} \epsilon_{\alpha_I}$ of one-particle energies as entries. The
interaction term ${\cal H}_{\rm int}$ yields a matrix with entries
\begin{equation} \label{interaction1}
<\!\alpha'_1 \ldots \alpha'_N | {\cal H}_{\rm int}| \alpha_1 \ldots \alpha_N\!>
= \!\!\sum_{IJ}\!\! \left( \prod_{i\neq IJ} \delta_{\alpha'_i \alpha_i}\!\!
\right)
Q_{\alpha'_I \alpha'_J \alpha_I \alpha_J}
\end{equation}
where
\begin{equation}
\label{interaction2}
Q_{\alpha'_I \alpha'_J \alpha_I \alpha_J} \equiv \sum_{p,p'=1}^M
\psi^*_{\alpha'_I}(p') \psi^*_{\alpha'_J}(p') u_{pp'} \psi_{\alpha_I}(p)
\psi_{\alpha_J}(p),
\end{equation}
$\psi_{\alpha_I}(p)$ denoting the amplitude of the wave-function in the
one-particle eigenstate $|\alpha_I >$ at the site $p$.

  The existence of the interaction yields two effects that we consider
separately. The diagonal matrix elements of  ${\cal H}_{\rm int}$ shift the
location of the $N$-body levels, an effect which is predominant for small
system size and large strength of the interaction, and which can yield an
important re-arrangement of the spectrum (see section~\ref{numeric_Ular}). This
situation is shortly described in the following subsection, and has been 
extensively discussed by Kamimura~\cite{Kamimura}, in the case of Anderson 
insulators with a very small localization length. The off-diagonal matrix 
elements of ${\cal H}_{\rm int}$ give rise to hopping among certain Fock 
states, and thus to delocalization in the Fock basis. In this study, we mainly 
focus on the description of this interaction induced delocalization in the Fock
basis, in the limit where the second effect dominates the first. This 
delocalization in the Fock basis is a generic effect of the interaction which 
should not be confused with delocalization in real space. It is only when the 
one-particle states are themselves localized in real space that Hilbert space 
delocalization may result in delocalization in real space. A recent 
analysis\footnote{In Ref.~\cite{ap}, it is also shown that the relation 
between the spectral rigidity and the level curvature is not direct for the 
$N$ body problem. For $N=1$, the original definition of the Thouless number, 
in terms of the curvature of the energy levels, coincides with our definition, 
based on the spectral rigidity. For $N=2$, these two definitions are not the 
same in the metallic regime, but the (spectral) Thouless number can be 
expressed in terms of another (topological) curvature, assuming that distinct 
Aharonov-bohm fluxes can be associated to the distinct particles. For $N\ge 3$,
the study of those different curvatures is postponed to a further 
study.}~\cite{ap} of the sensitivity of the energy levels to a change of 
boundary conditions has stressed this difference.

\subsection{Diagonal matrix elements of the interaction and large $U$-limit}

For very large $U$ and small system size $L$, the previously defined Fock basis
is no longer appropriate. It is more instructive to consider the Fock basis 
built of the on-site orbitals, and not of the one-particle eigenstates. The 
kinetic energy, and not the interaction, can then be treated perturbatively. In
this basis the $N$-particle states without kinetic terms can be classified 
according to the number of next-neighbor configurations, for a next-neighbor 
interaction. This limit will be discussed in more detail in 
section~\ref{numeric_Ular}, where numerical results show that at $U\approx 15$,
the spectrum of three spinless Fermions is split into three separated bands, 
with a density of states approximately given by the sum of three Gaussians 
centered at $E=0$, $U$, and $2U$. The weights of those three bands are directly
related to the number of next-neighbor configurations in the on-site Fock basis
states.

  However, for $U\approx 1$ we are far from seeing interaction induced gaps in 
the spectrum. We then assume that the diagonal matrix elements of the
Hamiltonian in the Fock basis built of the one-particle eigenstates, are mainly
dominated by the one-particle contributions, and that the effects coming from
the interaction can be neglected for those elements. Therefore, we consider 
only the delocalization in this basis, which results from the
{\it off-diagonal} terms.

\subsection{Off-diagonal matrix elements of the interaction}

  For $N \geq 3$, one can see from Eq.~(\ref{interaction1}) that there are only
non-zero matrix elements between Fock states having $N-2$ one-particle states
in common. This means that the Hamiltonian (\ref{hamiltonian2}) is a 
{\it sparse} matrix in this Fock basis, with a strongly preferential basis. 
This property was not present in the former studies\cite{shepelyansky,imry,wp} 
for $N=2$, and a straightforward generalization of the former results to $N$ 
particles would require $N$-body interactions. The two-body form of the 
interaction introduces specific problems for $N \geq 3$,  which has been 
recently discussed in Ref.~\cite{flambaum,shepelyansky-sushkov}. Moreover, 
there is a large degeneracy of these non-zero terms. For instance, when $N=3$, 
all the elements 
$<\alpha_1\alpha_2\beta|{\cal H}_{\rm int}|\alpha'_1\alpha'_2\beta'>=
Q_{\alpha_1\alpha_2\alpha'_1\alpha'_2}\delta_{\beta\beta'}$ 
are the same for all of the one-particle states $\beta=\beta'$.

 The form of the distribution of the degenerate non-zero off-diagonal terms
(Eq.\ref{interaction2}) is by itself a non-trivial one-particle problem. If the
underlying classical one-particle dynamics is diffusive, as in a disordered 
metal, ballistic chaotic, as in a billiard, or integrable, one gets different 
estimates~\cite{agkl,blanter} for the magnitude of the interaction matrix 
elements. For simplicity we will use the very rough approximation of 
uncorrelated one-particle wave-functions, with amplitude of the order 
$1/\sqrt{M}$ on each site with a random sign. This corresponds to a
one-particle Hamiltonian being statistically invariant under orthogonal
transformations ($O(M)$ invariance assumed in standard Random Matrix Theory).
As pointed out in Ref.~\cite{agkl,blanter}, this evaluation of the interaction
matrix elements only reproduces\footnote{We note two important modifications to
our rough approximation for the matrix elements. For energy transfer smaller 
than the Thouless energy $E_c$, its magnitude is enhanced to the order of 
$\pm \Delta_1/g_1$ when $g_1 \geq 1$. In one dimension ($g_1 \leq 1$), it was 
pointed out by Ponomarev and Silvestrov~\cite{ponomarev} that there are 
important modifications of the matrix elements. More precisely, as recently 
shown in Ref.\cite{waintal}, when the one particle states are localized, the 
fluctuations of the interaction matrix elements are so large that the effective
density of directly coupled Fock states becomes multifractal.} the zero wave 
mode contribution of a diffusion process. This gives 
$Q_{\alpha'_1 \alpha'_2 \alpha_1\alpha_2}\approx \pm Q_{\rm typ} \approx \pm 
U/M^{3/2}$ for on-site interactions. We use this approximation for the 
simplified theoretical picture that we present before the numerical study. This
is because we want to compare our predictions to simulations on disordered 
systems with too small sizes to have one-particle diffusion. The more detailed 
description of the off-diagonal matrix element will modify the quantitative 
dependence as a function of the system parameters, but will not change the 
general scheme of the effect of the interaction on the spectral correlations.  

Moreover, we will neglect the energy dependence of the $N$-particle density
$\rho_N$, taking $\rho_N^{-1}=\Delta_N \approx B_1/M^N$. $\Delta_N$ denotes the
typical $N$-particle level spacing in the bulk of the $N$-particle spectrum and
$B_1$ is the bandwidth (one-particle kinetic energy scale). 

\section{Spread width over states directly coupled by the interaction}

For $N=2$, the full Hamiltonian can be modeled \cite{wp} by a Gaussian ensemble
of random matrices with a preferential basis \cite{ps}. The structure of the 
projections $C_{\beta n}\equiv < \Psi_{\beta}|A_n >$ of the many-body 
eigenfunctions $|\Psi_{\beta}> $ (labeled by $\beta=1,\dots,n_{\rm tot}$) onto 
the Fock states $|A_n >$ is well described by the Breit-Wigner form 
\cite{wp,jacquod}
\begin{equation}\label{breit-wigner}
\langle |C_{\beta n}|^2 \rangle = \Delta_2
{\Gamma_2\over {2\pi [(E_{\beta}-E_{n})^2+\Gamma_2^2/4]}}\, ,
\end{equation}
where the brackets denote ensemble averaging, and the spread width
\begin{equation}\label{fermis_golden_rule_2}
\Gamma_2 = 2 \pi \langle Q^2\rangle \rho_2 \approx 2 \pi {U^2 \over
M^{3}} {M^2 \over B_1 }
\end{equation}
increases with the interaction according to Fermi's golden rule. This means
that (for $\Gamma_2 > \Delta_2$) an eigenfunction $|\Psi_{\beta}> =\sum_n 
C_{\beta n} |A_n >$ has significant projections on typically 
$\Gamma_2/\Delta_2$ Fock states.

 For $N \geq 3$, this can be generalized to a spreading width 
$\Gamma_N^{\rm (d)}\propto \langle Q^2 \rangle \rho_N^{\rm (d)}$ where
$\rho_N^{\rm (d)}$ is the density of the $N$-body Fock states {\it directly}
coupled by the interaction. For spinless Fermions, one has
$n_{\rm tot}=M!/(N!(M-N)!)$ Fock states and the number of Fock states directly
coupled by the two-body interaction is $n_{\rm eff}^{\rm (d)}=
N(M-N)+N(N-1)(M-N)(M-N-1)/4$. In the dilute limit $N\ll M$, one finds
$n_{\rm tot}\propto M^N/N!$ while $n_{\rm eff}^{\rm (d)} \propto  M^2$ is much 
smaller than $n_{\rm tot}$. Assuming uniform densities, this means that the 
effective level spacing $\Delta_N^{\rm (d)}$ between Fock states directly 
coupled by the interaction is of order $\Delta_2^{\rm eff} \approx 
B_1/n_{\rm eff}^{\rm (d)}$. For very few particles, $\Delta_2^{\rm eff} 
\approx \Delta_2$, but we emphasize that this approximate relation becomes 
uncorrect if $N$ is large, mainly in the finite density case where 
$N \propto M$. 

 A quantity closely related to the local density of states is the participation
ratio $R = \langle \sum_{n=1}^{n_{\rm tot}} |C_{\beta n}|^4 \rangle^{-1}$,
which gives the number of Fock states mixed by the interaction. Using the
structure (\ref{breit-wigner}) of the eigenfunctions at $\Gamma_2>\Delta_2$,
one can get the estimate $R \sim \pi \Gamma_N^{\rm (d)}/\Delta_N^{\rm (d)}
\approx 2\pi^2 U^2 (M^{3} B_1^2)^{-1} (n_{\rm eff}^{\rm (d)})^2$. Therefore we
expect to find $R \propto U^2$, since the contribution of the states directly
coupled by the interaction will dominate for small $U$.

  Therefore, the first observable effect of the interaction will be the
broadening of a Fock state over $\Gamma_N^{\rm (d)}/ \Delta_N^{\rm (d)}$ other 
Fock states separated by a characteristic scale $\Delta_N^{\rm (d)} \approx 
B_1 / n_{\rm eff}^{\rm (d)} \approx \Delta_2^{\rm eff}$. This spreading width 
is proportional to $U^2$, but does {\it not} characterize the coupling of the 
original Fock State to the $N$-body spectrum. In this spread width, there are 
many Fock states (of a density $\rho_3 = 1/\Delta_3$ for $N=3$) which are not 
directly coupled to the original Fock state at this order in $U$. This is the 
major difference between $N=2$ and $N\geq 3$: For $N=2$, the width $\Gamma_2$ 
which characterizes the local density of interacting states in the Fock basis 
is directly related to the spectral statistics: the energy scale $E_{U}$ up to 
which the spectrum exhibits the universal Wigner-Dyson rigidity is 
given~\cite{wp} by this spread width $\Gamma_2$, provided $\Gamma_2 > 
\Delta_2$. For $N \geq 3$, even when $\Gamma_N^{\rm (d)} > \Delta_N$, the 
levels separated by $\Delta_N$ are not necessarily coupled and can be 
statistically independent. For the level repulsion at the scale $\Delta_N$, the
spreading width $\Gamma_N^{\rm (d)}$ does not provide a relevant energy scale. 
The $N$-particle Thouless number $g_N$ is not given by 
$\Gamma_N^{\rm (d)}/ \Delta_N^{\rm (d)}$, when $N \geq 3$.

  In the following, we discuss what should provide this relevant energy scale
for the spectral statistics of {\it consecutive} $N$-body levels, with a
density $\rho_N = 1/\Delta_N$, and thus the relevant $g_N$. We consider first
the case $N=3$. The generalization to an arbitrary number $N$ of particles is
straightforward, as far as we are in the dilute limit.

\section{ Perturbative regime ($ U \leq U_{c1}$) }

  The spectrum without interaction contains only one-particle correlations
appearing on energy scales larger than $\Delta_1$ and is close to uncorrelated
levels on lower energy scales. The interaction re-establishes~\cite{Moriond}
the universal Wigner-Dyson rigidity up to an energy $E_{U}$ which depends on 
the strength of the interaction. When $U$ is weak, it can couple only 
quasi-degenerate states and leads at most to Rabi-oscillations between 
adjacent levels.

  For $N=2$, $E_{U} \propto |U|$, while one finds $ E_{U} = \Gamma_2 \propto 
U^2$ at larger $U$ when $g_2 \equiv \Gamma_2 / \Delta_2 >1 $, i.e. when many 
Fock states are coupled and Fermi's golden rule applies. This can be
understood~\cite{wp} from the following arguments. For very weak $U$ 
($\Gamma_2 < \Delta_2$) only the coupling between two Fock states with a 
separation $\leq \Delta_2$ is relevant. This restricts the problem to the 
analysis of a solvable $2 \times 2$ random matrix, with diagonal terms 
typically much larger than the off-diagonal coupling term. A model with 
independent Gaussian entries, the variance of the diagonal entries being much 
larger than these of the off-diagonal term, was exactly solved in 
Ref.\cite{ps}. Since a $2 \times 2$ real symmetric matrix can be diagonalized 
by a rotation of angle $\theta$, it is easy to write the probability 
distribution in terms of the two eigenvalues and of $\theta$. Integrating over 
$\theta$ gives the joint probability distribution of the two eigenvalues. One 
finds for this $2 \times 2$ matrix model that $E_U$ is given by the absolute 
value (r.m.s) of the off-diagonal term.  It was interpreted in terms of Rabi 
oscillations between two Fock states at typically $\Delta_2$ away from each 
other in energy, and the range $E_{U}$ of the level repulsion was 
identified\footnote{A related effect is 
known~\cite{anderson-lee,montambaux,casati} for the one-particle problem, 
where the Thouless sensitivity to boundary conditions is proportional to the 
square root of the Landauer conductance when $g_1<1$ and to the Landauer 
conductance when $g_1 > 1$.} with this Rabi frequency. For energy separation 
$\epsilon < E_U$, the consecutive levels repel each other as in standard random
matrix theory, while their fluctuations are uncorrelated for $\epsilon > E_U$.

 For $N=3$, we denote by $|A_n> = |\alpha_1\alpha_2\alpha_3>$ the 
three-particle Fock states, of energy $E_{n} =
\epsilon_{\alpha_1}+\epsilon_{\alpha_2}+\epsilon_{\alpha_3}$, and we consider 
two energetically nearby Fock states $\mid A_n >$ and $\mid A_{n'} >$: 
i.e.~with $E_{n} - E_{n'} \approx  \Delta_3$, the three-particle level spacing.
For a weak interaction, as pointed out by Shepelyansky and Sushkov 
\cite{shepelyansky-sushkov}, the {\it effective} matrix element $U_3^{\rm eff}$
of the interaction between those two consecutive Fock states can be estimated 
using perturbation theory. It is only in {\it second order} that one gets a
non-zero contribution resulting from terms like
\begin{equation}
\sum_{n''}{{ <A_{n} | U_{12} |A_{n''} > < A_{n''} | U_{23} | A_{n'}>}
\over {E_{n} - E_{n''}}}\, ,
\end{equation}
where particle 1 interacts with particle 2, then particle 2 with particle 3.
Since we have a two-body interaction ($ <A_{n} | U_{12} | A_n''> =
< \alpha_1 \alpha_2 | U_{12} | \alpha''_1 \alpha''_2 >
\delta_{\alpha_3\alpha''_3}$), the summation over $n''$ is reduced to a sum
over the single-particle quantum number $\alpha''_2$. This sum  is of the order
of its largest term, i.e.~of a term with an energy denominator of order 
$\Delta_1$, the one-particle level spacing, and not $\Delta_3$. This eventually
gives for the effective matrix element which couples two consecutive Fock 
states a magnitude of order
\begin{equation}
U_3^{\rm eff} \approx \pm {U_{\rm typ}^2 \over \Delta_1} \approx 
\pm  {{ U^2} \over {M^3 \Delta_1}}\, 
\end{equation}
if one takes for $U_{\rm typ}$ our simple estimate \\
$Q_{\rm typ}= \pm U/M^{3/2}$.

 Therefore, in this perturbative regime, Fock states at an energy
$\Delta_3^{\rm (d)}$ from each other are coupled by a matrix element
$U_3^{\rm (d)} \approx U_{\rm typ}\approx \pm U/M^{3/2}$ while Fock states at 
$\Delta_3$ from each other are only coupled by $U_3^{\rm eff} \propto \pm 
U_{\rm typ}^2 / \Delta_1 \approx \pm U^2/(M^3 \Delta_1)$. From this, we draw 
two main conclusions for the three particle problem that we extend to an 
arbitrary number $N$ of particles.

\subsection{Hierarchical structure in the Fock basis}

\begin{figure}[tb]
\centerline{\epsfxsize=0.95\colwi\epsffile{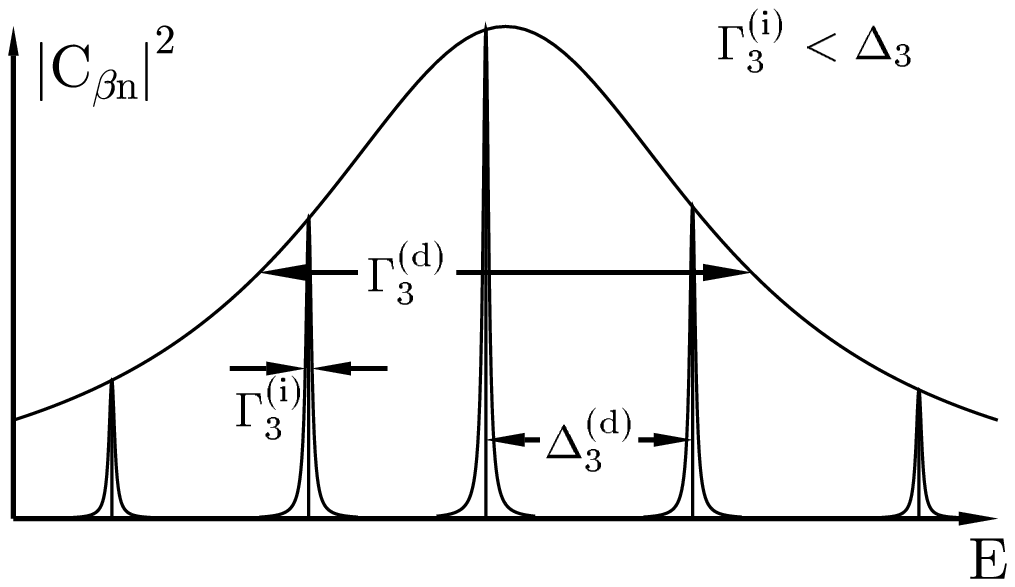}}
\caption[schema1]{\label{schema1}
Structure of the eigenfunction at $U<U_{c1}$ in the Fock basis}
\end{figure}

 The states in the Fock basis have a very particular hierarchical structure, as
sketched in figure \ref{schema1}. A Fock state is broadened over a density 
$\rho_3^{\rm (d)} = 1/\Delta_3^{\rm (d)} \approx 1/\Delta_2^{eff}$ of 
neighboring Fock states. This broadening has a Breit-Wigner form characterized 
by a width $\Gamma_3^{\rm (d)} \approx U^2/(M^3 \Delta_3^{\rm (d)})$. The 
projections over the neighboring Fock states at $\Delta_3^{\rm (d)}$ away from 
each other are themselves
 
\noindent
broadened over a density $\rho_3=\Delta_3^{-1}$, 
with a Breit-Wigner shape characterized by a width $\Gamma_3^{\rm (i)} \approx 
(U_3^{\rm eff})^2 \Delta_3^{-1}$. In the perturbative regime, 
$\Gamma_3^{\rm (i)} \le \Delta_3$ when $U < U_{c1}$.

\subsection{Rabi frequency and Wigner-Dyson rigidity}

 When $U$ is so small that the broadening $\Gamma_3^{\rm (i)}$ is smaller than
$\Delta_3$, the levels are essentially uncorrelated. However, level repulsion 
occurs for energy scales smaller than $E_U \approx |U_3^{\rm eff}| < \Delta_3$.
The reason for this is a straightforward generalization of the argument given 
for $N=2$: The energy level correlations come from the very small coupling 
terms between Fock states which are nearest neighbors in energy. This reduces 
the complicated problem of a very large random matrix with a few non-zero small
off-diagonal terms to the solvable problem of a $2 \times 2$ matrix with an 
{\it effective} off-diagonal term of magnitude $U_3^{\rm eff}$. 
Rabi-oscillations between the two coupled diagonal terms occur. Their 
frequency, of order $|U_3^{\rm eff}|$, characterizes also the scale $E_{U}$ 
below which the universal level repulsion occurs. For $N=2$, this gives $E_U 
\propto |U|$ (direct coupling), while $E_U \approx |U_3^{\rm eff}| \propto 
|U^2|$ for $N=3$.

 Let us consider now the case where $N=4$. As for the three particle case, it 
is sufficient to go to the second order in perturbation theory for having a non
zero effective matrix element coupling two consecutive Fock states. One has:
\begin{equation}
U_4^{\rm eff} \approx 
\sum_{n''}{{ <A_{n} | U_{12} |A_{n''} > < A_{n''} | U_{34} | A_{n'}>}
\over {E_{n} - E_{n''}}}, 
\end{equation}
where the $|A_{n}>$ are now the Fock states for 4 particles. The difference 
with the case where $N=3$ is that the summation over $n''$ is now totally 
suppressed. This yields a smallest possible denominator $ E_{n} - E_{n''} $ of 
order $B_1$ and not $\Delta_1$. One finds $ U_4^{\rm eff} \approx \pm 
U_{\rm typ}^2 /B_1$. Similarly, one finds $U_5^{\rm eff} \approx \pm  
U_{\rm typ}^3 / (B_1 \Delta_1)$, $U_6^{\rm eff} \approx \pm 
U_{\rm typ}^3/ (B_1^2)$... and the general expression is given by 
$U_N^{\rm eff} \approx \pm  B_1 (U_{\rm typ}/B_1)^P $ for an even number $N$ of
particles with $P=N/2$, and  $U_N^{\rm eff} \approx \pm 
(U_{\rm typ}/ B_1)^P (B_1^2/\Delta_1)$ for an odd number $N$ with $P=(N+1)/2$. 
This gives us the energy range $E_U \approx |U_N^{\rm eff}|$ of the level 
repulsion for weak interaction and arbitrary $N$.  

\section{Crossover from Poisson to Wigner: $U=U_{c1}$.}

 When $\Gamma_3^{\rm (i)} \approx (U_3^{\rm eff})^2/ \Delta_3 \approx 
\Delta_3$, the Wigner-Dyson rigidity is established on the scale $\Delta_3$ for
the three particle case. This defines the first interaction threshold $U_{c1}$ 
where a sharp cross-over from Poisson to Wigner should be observed in the 
distribution $P(S)$ of the spacings between consecutive energy levels. For an 
arbitrary number $N$ of particles, the first threshold $U_{c1}$ is given by the
condition
\begin{equation}\label{gnud}
g_N  \approx {|U_N^{\rm eff}| \over \Delta_N} \approx 1 \, .
\end{equation}
 Using the estimate for $U_N^{\rm eff}$ given by the first non-zero order in 
the perturbative expansion in $U$, and assuming that $N$ is small enough to 
have $\Delta_2 \approx \Delta_2^{\rm eff}$, (i.e. neglecting a factor of order
$1/N^2$), one finds the general relation 
\begin{equation}\label{gng2}
g_N \approx g_2^P \, .
\end{equation}
where $P=N/2$ when $N$ is even and $P=(N+1)/2$ when $N$ is odd. Therefore, the 
interaction $U_{c1}$ where the two particle Thouless number $g_2$, given by 
$(U/B_1)^2 M$ in our estimate, is of order one (i.e.~$U_{\rm typ} \approx 
\Delta_2^{\rm eff}$), does not signify a Poisson-Wigner cross-over for the two 
particle case only, but also for the $N$-body spectrum, as far as we stay in 
the dilute limit. We note that our general relation\footnote{In 
Eq.~(\ref{gnud}), the Thouless numbers are defined by the energy scale below 
which level repulsion occurs, in units of the mean level spacing $\Delta_N$. 
This definition, valid in the perturbative regime only, differs from the 
definition used in Ref.\cite{shepelyansky-sushkov} ($g_N \equiv 
\Gamma_N^{\rm (i)} / \Delta_N \approx (U_N^{\rm eff} / \Delta_N)^2$). However, 
this subtlety (see the previous footnote) does not matter for 
Eq.~(\ref{gng2}).} implies $g_3 \approx g_2^2$, in agreement with 
Ref.~\cite{shepelyansky-sushkov}. 

  $U_{c1}$ can also be understood from the parametric motion of the energy 
levels when $U$ increases. Let us consider again the case where $N=3$. Level 
repulsion starts to be efficient when a certain relative characteristic energy 
change $\Sigma_{\rm R}^{\rm (3)}(U)$ of the levels due to the interaction is of
the order of the mean level spacing. The displacement resulting from the 
diagonal matrix elements of the interaction induces a parallel motion of the 
levels and is therefore irrelevant. For the real part of the self-energy 
$\Sigma_{\rm R}(U)$ of a given Fock state, perturbation theory gives different 
terms involving loops starting from the considered Fock State and visiting one,
two, three and more Fock states being directly coupled among each other by the 
two-body interaction before returning to the starting point. The term of order 
$U^2$ (loop visiting a single Fock state) cannot be relevant since it involves 
an energy denominator of order $\Delta_2^{\rm eff}$, and not $\Delta_3$. The 
first term involving loops visiting Fock states at $\Delta_3$ away from the 
unperturbed level is only given by the term of order $U^4$ (loop starting from 
the Fock state and visiting three different Fock states before return). For the
real part, this gives terms like
\begin{eqnarray}
\sum_{n' n'' n'''}&& { <A_{n} | U_{12} | A_{n'} > <A_{n'} | U_{23} | A_{n''} >
\over (E_{n}-E_{n'}) } \nonumber \\
&\times & { <A_{n''} | U_{23} | A_{n'''}> <A_{n'''}| U_{12} | A_{n}>  \over
(E_{n} - E_{n''}) (E_{n}-E_{n'''})}
\end{eqnarray}
 These sums are of the order of the term having the smallest possible
denominator, e.~g.\ $(E_{n}-E_{n''})\approx \Delta_3$, which {\it fixes} the 
state $\mid A_{n''}>$ and thus suppresses the summation over $n''$. Then, the 
two-body character of the interaction yields for the energy differences 
$E_{n} - E_{n'}$ and $E_{n} - E_{n'''}$, with $E_{n}$ and $E_{n''}$ fixed, a 
smallest possible value of order $\Delta_1$. One can see that 
$\Sigma_{\rm R}^{\rm (3)}(U) \approx \Delta_3$ precisely when 
$g_3=\Gamma_3^{\rm (i)} / \Delta_3 \approx 1$, for $U=U_{c1}$. It is 
straightforward to check that it is also at $U=U_{c1}$ that the term of order 
$U^4$ of the perturbative expansion of the imaginary part of the self-energy is
of order $\Delta_3$. Moreover, this argument can be easily extended to an 
arbitrary number $N$ of particles, with a conclusion in agreement with these 
previously presented in order to obtain $U_{c1}$. 

\section{Breakdown of the perturbation theory and non-ergodic
Wigner-Dyson regime: $U_{c1} < U < U_{c2}$}

 The argument that we propose to characterize the relevant Thouless numbers
$g_N$ is reminiscent of a locator expansion~\cite{Anderson58,Imry94} ``\`a la
Anderson'', where the self-energy of the Fock states in the presence of the
hopping terms yielded by the interaction is evaluated using perturbation
theory. For one-particle localization, the breakdown of this perturbative
expansion in the basis built of the site orbitals was related to a
metal-insulator transition due to delocalization in real space. One knows from 
the scaling theory too that this transition occurs for $g_1 \approx 1$. This is
a closely related consideration which has led the authors of Ref.~\cite{agkl} 
to conjecture that interactions should give an Anderson transition in Fock 
space, for a critical value $\epsilon^{*}$ of the excitation energy of the 
extra quasi-particle injected above the Fermi sea. This led us to determine up 
to what maximum value of $U$ the relevant Thouless numbers $g_N$ can be given 
by perturbation theory. If one follows Shepelyansky and 
Sushkov\cite{shepelyansky-sushkov}, who assume that perturbation theory 
remains valid when $g_3 \geq 1$ for $N=3$, the relation $g_3 \approx g_2^2$ 
gives $g_3 = \Gamma_3^{\rm (i)}/\Delta_3 \propto U^4$ above $U_{c1}$. However, 
in our numerical study (see section \ref{numerics}) we observe above $U_{c1}$ a
$|U|$-increase of $g_3$, following the perturbative $U^2$-increase (Rabi 
regime). Moreover, all the quantities calculated for three spinless Fermions 
(local density of states, participation ratio, spectral rigidity) have not 
given any trace of a $U^4$-proportional behavior when $U > U_{c1}$. 

 This leads us to suspect that perturbation theory can only be used up to
$U_{c1}$, where level repulsion is established at the scale $\Delta_3$. Above
this threshold, the relevant $\Gamma_3$ does not coincide with the perturbative
estimate $\Gamma_3^{\rm (i)}$. This breakdown of the perturbation theory above 
$U_{c1}$, when $U_{\rm typ} > \Delta_2^{\rm eff}$, can be shown if one 
evaluates the effective matrix element coupling nearby Fock states at higher 
orders in $U$. Let us present the argument for $N=4$. The coupling term of 
order $U^2$ was found to be $U_4^{\rm eff} \approx \pm U_{\rm typ}^2 / B_1$. 
This corresponds to a process where particle 1 interacts with particle 2, then 
particle 3 with particle 4. A term of order $U^3$ is given for instance if 
particle 3 interacts with particle 4 once more. 
\begin{eqnarray}
&&U_4^{\rm eff} ({\rm order\ 3}) \approx \\
&&\sum_{n'',n'''}\!\!
{{<\! A_n|U_{12}|A_{n''}\! ><\! A_{n''}|U_{34}|A_{n'''}\! >
<\! A_{n'''}|U_{34}|A_{n'}\! >} \over 
{(E_n-E_{n''})(E_n-E_{n'''})}} \nonumber
\end{eqnarray}
 The smallest energy denominator is the product of two energy differences. The 
first one is of order $B_1$, but one has extra degrees of freedom for the 
choice of two one particle quantum numbers of $|A_{n'''}>$ since particle 3 
interacts with particle 4 two times. This reduces the second energy difference
to a smallest value of order $\Delta_2^{\rm eff}$, and gives $U_4^{\rm eff} 
({\rm order\ 3}) \approx U_4^{\rm eff} (U_{\rm typ} /\Delta_2^{\rm eff})$. 
This means that the terms in $U^2$ and in $U^3$  of the perturbative expansion 
are of the same order for $U_{\rm typ} \approx \Delta_2^{\rm eff}$. The 
generalization is straigthforward: For arbitary $N$, one finds
\begin{equation}
U_N^{\rm eff} ({\rm order}\ p) \approx U_N^{\rm eff} 
\left( {U_{\rm typ} \over \Delta_2^{\rm eff}}\right)^{p-P} 
\end{equation} 
for the terms of order $p > P$, where $P$ is the order giving $U_N^{\rm eff}$.
This indicates that the sum to all orders in $U$ does not converge above 
$U_{c1}$.  Similar considerations, used here to evaluate to all orders the 
effective matrix element coupling Fock states which are nearby in energy, can 
be given for the self-energy. From this emerges the general result that the 
perturbative sum in $U$ has terms with similar magnitude to all orders when 
$U_{\rm typ} \approx \Delta_2^{\rm eff}$. One can then conclude that 
perturbation theory breaks down at the Poisson-Wigner cross-over in the 
$N$-body spectrum. If one compares our conclusions with the ones presented in 
Ref.\cite{shepelyansky-sushkov}, we also find that the threshold $U_{c1}$ 
corresponds to $g_2 \approx 1$ for very few particles. However, in contrast to 
the conclusions of Ref.\cite{shepelyansky-sushkov}, we have shown that the 
relation $g_3 \approx g_2^2$ for $N=3$ cannot be extended for $g_2 > 1$ and 
holds only when $g_2 \le 1$. For finite $N$ in a finite size system, the 
spectral statistics exhibits a cross-over from Poisson to Wigner. When $N$ 
increases, this cross-over should become sharper and sharper to eventually 
give a real transition. It is natural that such a transition is accompanied by 
a breakdown of perturbation theory of the self-energy of the Fock states, for 
$U_{\rm typ} \approx \Delta_2^{\rm eff}$, as it happens in the locator 
expansion of the self-energy for 

\begin{figure}[tb]
\centerline{\epsfxsize=0.95\colwi\epsffile{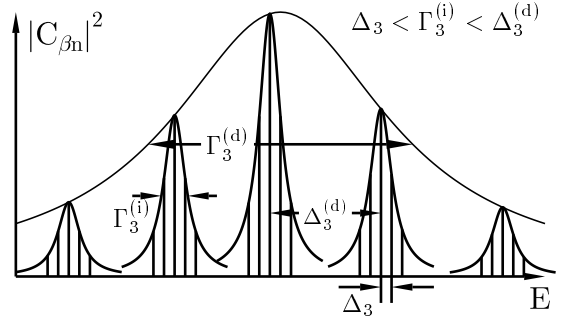}}
\caption[schema2]{\label{schema2}
Structure of the eigenfunction at $U_{c1}<U<U_{c2}$ in the Fock basis}
\end{figure}
\noindent
the one-particle problem at $g_1 \approx 1$. 
In both cases the condition is  that the matrix element of a state to the 
"first generation" of states directly coupled to it by the interaction, becomes
comparable to the level spacing of the latter states. A similar picture and 
delocalization condition applies also to the analogous transition in the 
Hilbert space of different numbers of excited quasiparticles in a degenerate 
Fermi system, suggested~\cite{agkl} recently by Altshuler et al..

 As mentioned above, $g_3$ does not increase as $U^4$ when $g_3 \gtrsim 1$ 
(see figure~\ref{EU23}). We are however not able to explain the observed linear
increase for $U>U_{c1}$. Nevertheless, one can say that the interacting states 
for $U>U_{c1}$ should not be ergodic in the energy window where they are 
broadened, but still have a structure as sketched in figure \ref{schema2}. 
When $\Gamma_3 \equiv g_3 \Delta_3$ is much smaller than $\Delta_3^{\rm (d)}
\approx \Delta_2^{\rm eff}$, there are still many Fock states inside the 
energy width $\Gamma_3^{\rm (d)}$ where an interacting state has essentially a 
zero projection. This will disappear only at a second threshold $U_{c2}$,
characterized by the condition: $\Gamma_3 (U_{c2})=\Delta_3^{\rm (d)}$.

\section{Ergodic Wigner-Dyson Regime and Self-consistent Theory: $U>U_{c2}$}

   We present here a conjecture for the regime of rather large interactions, 
when a Fock state is well coupled by the interaction to the three-body 
spectrum. We consider again the case where $N=3$ and a sufficiently strong 
interaction for having $\Gamma_3 \geq \Delta_3^{\rm (d)}$, but nevertheless 
small enough for not being in the large $U$-limit dominated by the diagonal 
terms of the interaction. For $U > U_{c2}$, one can assume that the 
interacting states are not unambiguously related to the previous hierarchy of 
Fock states, but are closer to  random mixtures of $g_3$ consecutive Fock 
states, each of them contributing with a projection of random sign and of 
typical amplitude of order $1/\sqrt{g_3}$. In other words, one has a simpler 
case where all the Fock states in an energy window $\Gamma_3$ are now well 
coupled, and remain decoupled from the other Fock states outside this window.
This is what we mean by ``ergodic Wigner-Dyson regime'', where the sparsity of 
the 3-body 

\begin{figure}[tb]
\centerline{\epsfxsize=0.95\colwi\epsffile{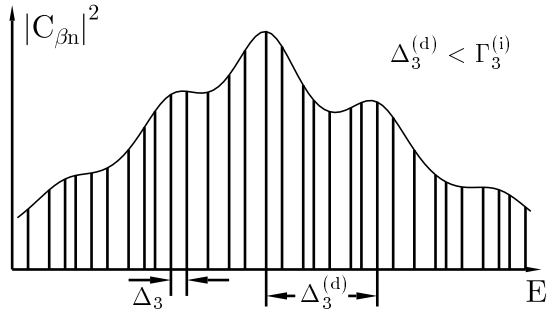}}
\caption[schema3]{\label{schema3}
Structure of the eigenfunction for $U>U_{c2}$ in the Fock basis}
\end{figure}
\noindent
Hamiltonian yielded by the pairwise character of the interaction 
becomes essentially irrelevant. A self-consistent evaluation of $g_3$ becomes 
possible if one assumes that an interacting state $|\Psi_{\beta}>$ has the 
following structure in the Fock basis:

\begin{equation}
|\Psi_\beta>=\sum_{\alpha_1\alpha_2\alpha_3} 
C^{\beta}_{\alpha_1\alpha_2\alpha_3}|\alpha_1\alpha_2\alpha_3>.
\end{equation}
 For $N=3$, we calculate the interaction matrix element $Q_{\beta'\beta}$
between two states $|\Psi_{\beta}>$ and $|\Psi_{\beta'}>$ at nearby energies, 
each of them being superpositions (with amplitude $C^{\beta}_{\alpha_1
\alpha_2 \alpha_3}\approx 1/\sqrt{g_3}$) of $g_3=\Gamma_3/\Delta_3$ Fock states
($\Delta_3=B_1/n_{\rm tot}$) for $L\leq L_1$:
\begin{equation}\label{self_element}
Q_{\beta'\beta}= \!\!
\sum_{\alpha'_1 \alpha'_2 \alpha'_3 \atop \alpha_1 \alpha_2 \alpha_3} \!
C^{\beta'}_{\alpha'_1 \alpha'_2 \alpha'_3} 
C^{\beta}_{\alpha_1 \alpha_2 \alpha_3}
 <\alpha'_1 \alpha'_2 \alpha'_3 | {\cal H}_{\rm int}
| \alpha_1 \alpha_2 \alpha_3> \,
\end{equation}
where the sums have to run over $g_3$ basis states. This means, that each of
the $\alpha $-summations runs over $g_{3}^{1/3}$ values. As can be seen from
(\ref{interaction1}), the matrix elements between the Fock states contain
three terms of the form
$Q_{\alpha'_I \alpha'_J \alpha_I \alpha_J}\delta_{\alpha'_K\alpha_K}$ where
$\{I,J,K\}$ are the different cyclic permutations of $\{1,2,3\}$. The
Kronecker-$\delta $ reduces the relevant summations occurring in
(\ref{self_element}) to 5 summations, each of them running over roughly
$g_{3}^{1/3}$ values. Thus, $Q_{\beta'\beta}$ consists of a sum of 
$3 g_3^{5/3}$ terms of typical size $Q_{\rm typ}/g_{3}$ with random sign, 
which yields the result
\begin{equation}
\langle |Q_{\beta' \beta }|^2 \rangle \approx 3 g_{3}^{5/3}
\frac{Q_{\rm typ}^2}{g_{3}^2}= \frac{3 U^2}{M^{3}g_{3}^{1/3}}\, .
\end{equation}
Plugging this into a Fermi golden rule evaluation of the spread width
\begin{equation}
\Gamma_3 =\frac{2 \pi \langle |Q_{\beta'\beta}|^2 \rangle }{\Delta_3}\, , 
\end{equation}
one obtains
\begin{equation}
g_{3}=\frac{\Gamma_3}{\Delta_3} = \frac{ (6 \pi)^{3/4}}{M^{9/4}}
\left(n_{\rm tot}\frac{U}{B_1}\right)^{3/2} .
\end{equation}
At $U>U_{c2}$, we therefore expect the decay width of the eigenfunctions and
the participation ratio $R$ to increase like $U^{3/2}$.

  For $N$ spinless Fermions, this ergodic Wigner Dyson regime is characterized 
by $g_N \propto U^{N/(N-1)}$, as it can be seen from a straightforward 
generalization of the self-consistent argument presented above for $N=3$.

\section{Numerical study of three spinless Fermions}\label{numerics}

  In order to illustrate our theory, we have performed a numerical study of the
many-body eigenstates and eigenenergies for three spinless Fermions in a 
disordered cubic lattice.

\subsection{Numerical model and characteristic scales}

 For the numerical simulations, we use a three dimensional tight binding model 
on a cubic lattice containing $3\times 3\times 3$ sites. The disorder and the
hopping terms are described by the usual Anderson Hamiltonian with on-site
potentials drawn from a rectangular distribution of width $2W$ with $W=2$ and
nearest neighbor hopping terms $V\equiv 1$ which set the energy scale. We use
rigid boundary conditions in all three directions. In addition, we use a
two-body interaction of the form (\ref{hamiltonian2}) with $u_{ij}= U$ when the
sites $i$ and $j$ are nearest neighbors on the lattice and $u_{ij}=0$
otherwise.

In such a cube, there are $M=27$ one-particle states with a typical density of 
$\rho_1=1/\Delta_1\approx 4$ in the center of the band. For spinless Fermions, 
this leads to  $M_2=M(M-1)/2=351$ two-particle states and 
$n_{\rm tot}=M_3=M(M-1)(M-2)/3!=2925$ three-particle states. For $U \approx 1$ 
and in the center of the band, the density of the three-particle levels is 
about $\rho_3=1/ \Delta_3  \approx 270$, while the density of two-particle 
levels amounts to $\rho_2=1/ \Delta_2 \approx 40$. The density of 
three-particle Slater determinants directly coupled to a given state by the 
interaction is larger: with the number 
$n_{\rm eff}^{\rm (d)}=N(M-N)+N(N-1)(M-N)(M-N-1)/4=900$ of non-zero 
off-diagonal interaction matrix elements in a line of ${\cal H}_{\rm {int}}$, 
one finds $\rho_3^{\rm (d)}=1/ \Delta_3^{\rm (d)}\equiv 1/ \Delta_2^{\rm eff}
\approx n_{\rm eff}^{\rm (d)}/(n_{\rm tot}\Delta_3)\approx 83$.

  For the analysis of the numerical results, we will slightly improve our 
estimates of the interaction matrix elements, taking into account that the 
interaction is not strictly on-site, but of range 1, since a particle on a 
given site can interact with another one when the latter is on one of the
adjacent sites. The matrix element coupling two states of the Fock basis 
(\ref{interaction2}) contains a double sum over the sites $p$ and $p'$ of the 
lattice. There are non-zero contributions to the sum whenever site $p'$ is a 
next neighbor of site $p$ on the lattice. In the cube we consider, there is one
site which has 6 next neighbors (NN), 6 sites with 5 NN, 12 sites with 4 NN and
8 sites with 3 NN. The mean number of next neighbors is $Z=4$ and the sums in
(\ref{interaction2}) run over a total number of $ZM$ terms. Assuming the
statistical invariance of the one-particle Hamiltonian under orthogonal 
transformations, this yields a typical size of the off-diagonal interaction 
matrix elements  $Q_{\rm typ} \approx \pm U\sqrt{Z}/M^{3/2}=\pm 0.014U$.

  In the same way, one finds $Q_{\alpha_1 \alpha_2 \alpha_1 \alpha_2}\approx 3 
Z U/M\approx 0.45 U$ for the diagonal terms of the interaction, which lead to a
shift of the diagonal elements of the Hamiltonian and conserve the sign of the 
interaction. The factor of three is due to the combinatorial factor which 
counts the number of different pairs out of three particles. For not too large 
$U$ we neglect them as compared to the fluctuations of the diagonal elements
(which are of the order of the band-width $B_1\approx 10$) for $U=0$.

  This must be carefully taken into account when estimating statistical 
properties of these matrices. In the Fermi golden rule formula for the spread 
width of the levels (\ref{fermis_golden_rule_2}) one must introduce the
effective level spacing $\Delta_3^{\rm (d)}$ of the directly available levels 
and finds the expression
\begin{equation}\label{fermis_golden_rule_n}
\Gamma_3^{\rm (d)} = 2 \pi \langle Q^2\rangle /\Delta_2^{\rm eff}
\approx 2 \pi U^2 Z M^{-3} \rho_3^{\rm (d)} \approx 0.11 U^2.
\end{equation}

  The effective interaction between basis states which are coupled by second
order processes only, can therefore be estimated to be
\begin{equation}
U_3^{\rm eff} \approx \pm \frac{\sqrt{3!} ZU^2}{M^3 \Delta_1}
\approx 0.002 U^2 \, ,
\end{equation}
leading to the spread width
\begin{equation}
\Gamma_3^{\rm (i)}\approx 2\pi \frac{|U_3^{\rm eff}|^2}
{\Delta_3}\approx 0.0067 U^4
\end{equation}
in the perturbative regime where $\Gamma_3^{\rm (i)} < \Delta_3$. From these 
estimates, we get the first threshold $U_{c1}\approx 0.85$ for which 
$U_{\rm typ} \approx \Delta_2^{\rm eff}$.

\subsection{Structure of the wave functions}

\begin{figure}[tb]
\centerline{\epsfxsize=0.95\colwi\epsffile{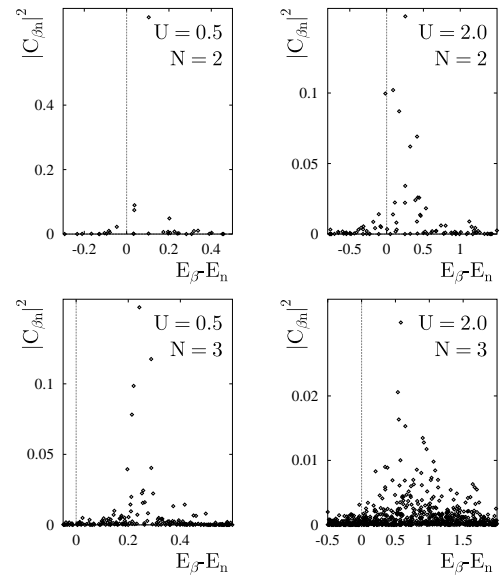}}
\caption[can]{\label{can}
The projections $|C_{\beta n}|^2$ of a typical eigenstate
$|\Psi_{\beta} > $ ($E_{\beta}\approx 0$) at interaction
strength $U=0.5$ (left) and $U=2.0$ (right) on the
basis of Fock states  $|A_{n}>$. The data are for
$N=2$ (upper) and $N=3$ (lower) spinless Fermions in a cube of 
$3\times 3\times 3$ sites
with interaction $U$ between next neighbors and on-site disorder $W=2$.}
\end{figure}

  We first concentrate on the structure of the eigenstates in the Fock basis. 
Examples are shown in Fig.~\ref{can}. In each of the pictures, only one 
eigenstate $|\Psi_{\beta} > $ at an energy $E_{\beta}\approx 0$ is shown. Each 
point represents the overlap $|C_{\beta n}|^2 = |< \Psi_{\beta} | A_{n} >|^2 $ 
with a Fock state $|A_{n} > $ and is plotted as a function of the energy 
difference $E_{\beta}-E_{n}$ between the eigenstate and the Fock state.

  It can be seen that in the case of two particles, almost all of the Fock 
states which are in a certain energy range around the energy of the unperturbed
eigenstate have a non-negligible overlap with it. For three particles, however,
many very small values of the projections onto Fock states occur, even when 
they are quite close in energy. However, it is difficult to observe the 
hierarchical structure of the three-particle states because the scales 

\noindent
$\Delta_3$ and $\Delta_2^{\rm eff}$ differ by a factor of 3 only in our case 
and because of the statistical fluctuations.

\begin{figure}[tb]
\centerline{\epsfxsize=0.75\colwi\epsffile{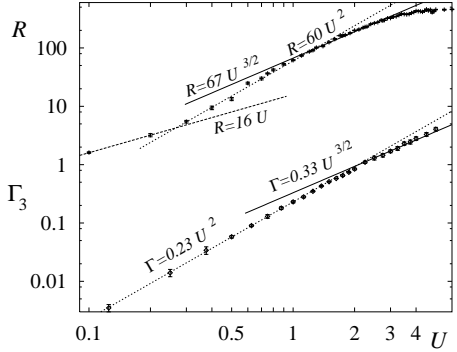}}
\caption[Ga_PR_3x3x3]{\label{Ga_PR_3x3x3}
Upper curve:
Participation ratio $R$ for 3 spinless Fermions in a
$3\times 3 \times 3$ disordered lattice ($V=1,W=2$) with
rigid boundary conditions and nearest neighbor interaction $U$.
The dashed, dotted and solid line represent power law fits in the different
regimes yielding $R = 16 U$, $60 U^2$, and $67 U^{3/2}$, respectively.
Lower curve:
Spread width $\Gamma_3$ characterizing the local density of states 
of 3 spinless interacting Fermions in the Fock basis. The dotted and 
solid line represent power law fits in the different regimes yielding 
$\Gamma_3 = 0.23 U^2$, and $0.33 U^{3/2}$, respectively.}
\end{figure}

  From these overlap matrix elements, taking into account several different 
realizations, we have computed the local density of interacting states in the 
Fock basis (Wigner strength function). In spite of the fact that there are many
very small values in the individual overlap matrix elements, the average is 
well described by the Breit--Wigner form (\ref{breit-wigner}). Its spread width
$\Gamma_3 $ is shown in Fig.~\ref{Ga_PR_3x3x3} (lower data points) as a 
function of $U$. Therefore by $\Gamma_3$ here, we mean the total spread width 
extracted from the average local density of interacting states in the Fock 
basis, and not the partial spread widths $\Gamma_3^{\rm (d)}$ and 
$\Gamma_3^{\rm (i)}$ introduced previously. $\Gamma_3 $ behaves quadratically 
down to rather low interaction values while $\Gamma_3 \propto U^{3/2}$ above 
$U\approx 1.5$.

For weak interaction, we are in the regime where the hierarchical structure of 
the eigenstates should be important. The spread width is then dominated by the 
spread width $\Gamma_3^{\rm (d)}$. Our estimates presented above give 
$\Gamma_3^{\rm (d)} \approx 0.11 U^2$ which is the correct order of magnitude.

From our theoretical considerations, we expect to obtain a regime in which the
wave-functions are ergodic and the sparseness of the Hamiltonian irrelevant 
when $U>U_{c2}\approx 1.8$. Taking into account the refined estimates of this
section, the spread width is expected to be given by the self-consistent 
expression
\begin{equation}
\Gamma_3=\Delta_3
\left(\frac{6 \pi Z U^2 }
{M^3 \Delta_3^2}\right)^{3/4} \approx 0.25 U^{3/2}\, .
\end{equation}

\noindent
In the numerical data, one observes indeed a transition at $U\sim 2$ to a 
regime in which $\Gamma_3 \propto U^{3/2}$ with a prefactor whose order of 
magnitude coincides again with the expected value.

 The upper points in Fig.~\ref{Ga_PR_3x3x3} show the participation ratio 
$R = \langle \sum_{n=1}^{n_{\rm tot}} |C_{\beta n}|^4 \rangle^{-1}$. The 
behavior of $R$ is quite similar to the one of the spread width $\Gamma_3 $. 
The participation ratio, which goes to $R=1$ at $U=0$, increases proportional 
to the square of the interaction in the regime $0.25 < U < 1.5$ and, as 
$\Gamma_3$, exhibits the signature of the ergodic regime for $U>1.5$. When 
calculating the participation ratio $R$, one has also to take into account, 
that not all of the states which are in the available energy interval can 
participate. Again, one has to replace the level spacing $\Delta_3$ by 
$\Delta_2^{\rm eff}$ to obtain $R \approx \pi \Gamma_3 /\Delta_2^{\rm eff} 
\approx 2\pi^2 U^2 Z (M^{3} (\Delta_2^{\rm eff})^2)^{-1}\approx 28 U^2$. In the
ergodic regime, the self-consistent estimate presented above gives $R\approx 
212 U^{3/2}$. However, at low $U$, a difference arises since $R=1$ and 
$\Gamma_{3}=0$ at $U=0$.

The ratio $R/\Gamma_3 \approx 260$ in the quadratic regime is much smaller than
the one expected from a democratic participation of the Fock states according 
to (\ref{breit-wigner}), which yielded $R/\Gamma_3 = \pi /\Delta_3 \approx 
850$. This is a consequence of the fact that individual states can show strong 
fluctuations around (\ref{breit-wigner}), thereby lowering the participation 
ratio. Furthermore, the sparse structure of the Hamiltonian, and the resulting 
hierarchical structure of the eigenfunctions reduces the number of 
participating basis states as has been seen in Fig.~\ref{can}.

\subsection{Spectral statistics}

\begin{figure}[tb]
\centerline{\epsfxsize=0.8\colwi\epsffile{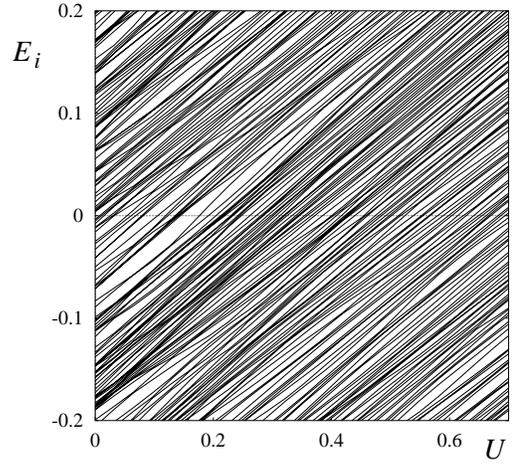}}
\caption[Spag_U]{\label{Spag_U}
Parametric dependence of the three-particle spectrum as a function of $U$,
around the band center. The change in spectral correlation can be seen.
For $U<0.2$, many levels seem to cross (weak level repulsion), while for 
larger $U$, the crossings are avoided (strong level repulsion).}
\end{figure}

The evolution of some energy levels in the center of the band as a function of 
$U$ is shown in Fig.~\ref{Spag_U}. First of all,

\begin{figure}[tb]
\centerline{\epsfxsize=0.7\colwi\epsffile{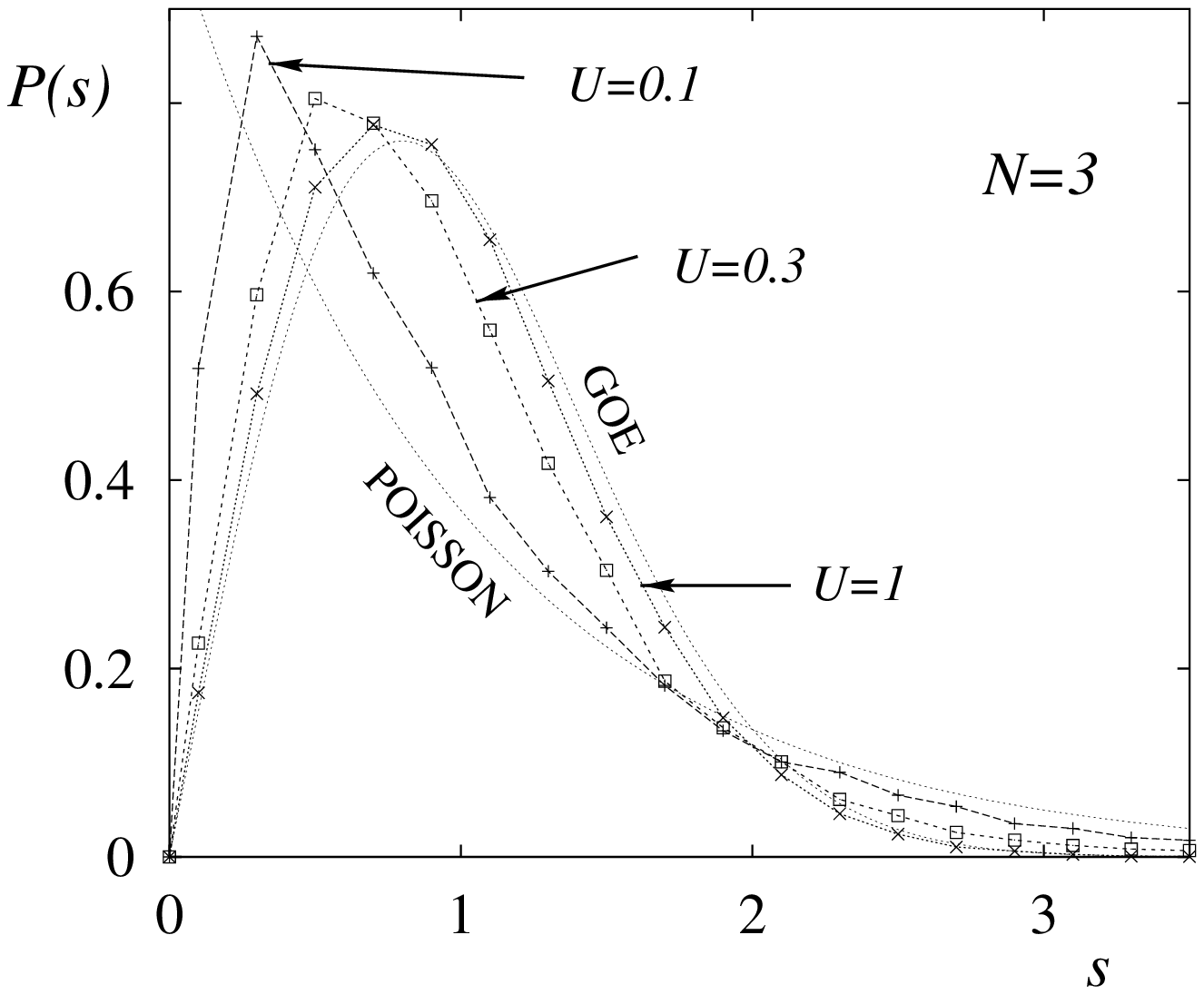}}
\centerline{\epsfxsize=0.7\colwi\epsffile{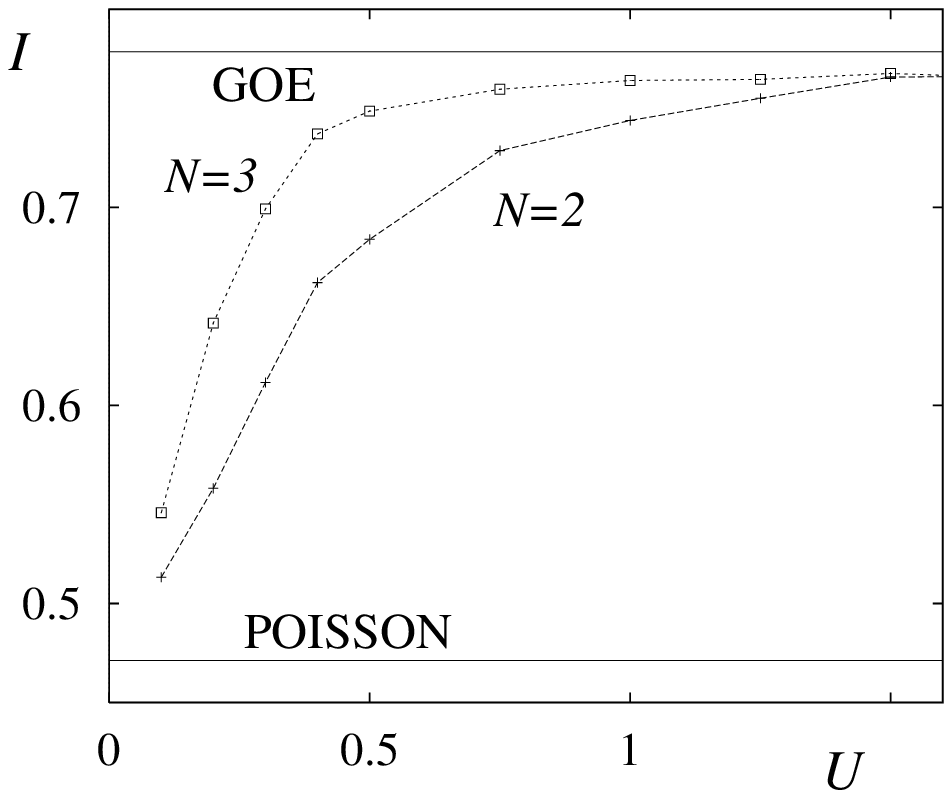}}
\caption[PS3]{\label{PS3}
Top:
Level spacing distribution $P(s)$ for three spinless Fermions
at $U=0.1$, $U=0.3$ and at $U=1$. For comparison, the Poisson and
GOE distributions are plotted also.
Bottom:
The integrated level spacing distribution $I$ as a function of the
interaction for two and three particles.}
\end{figure}
\noindent
the positive slope of all of 
the levels is visible. This is due to the diagonal matrix elements of the 
interaction which lead to a shift in energy of the order of $0.45U$ as expected
from the typical size of these elements.

But we can also observe changes in the statistical behavior of the spectrum. At
low $U$, there are strong fluctuations in the level spacing while the spectrum 
becomes 

\begin{figure}[tb]
\centerline{\epsfxsize=0.7\colwi\epsffile{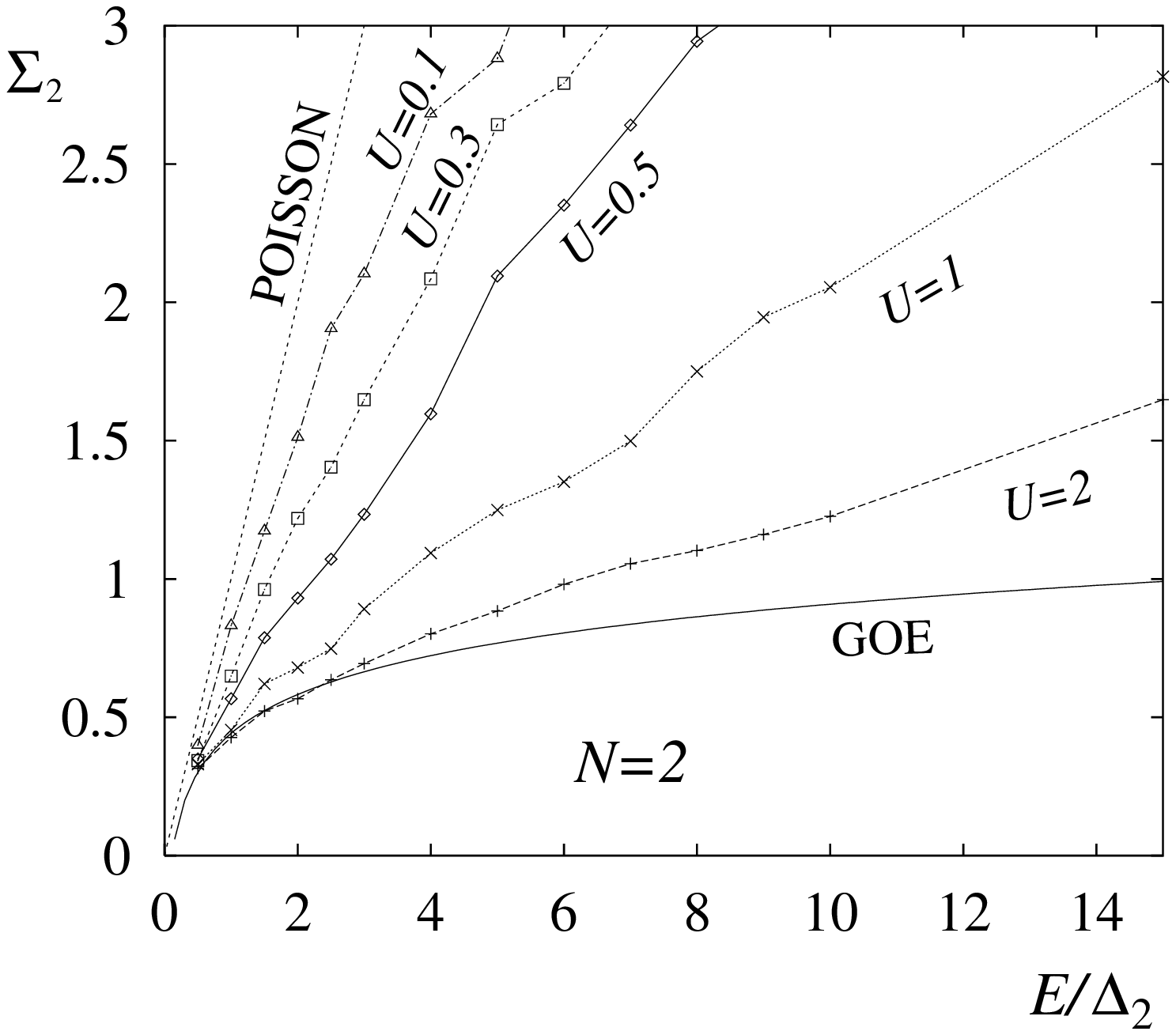}}
\centerline{\epsfxsize=0.7\colwi\epsffile{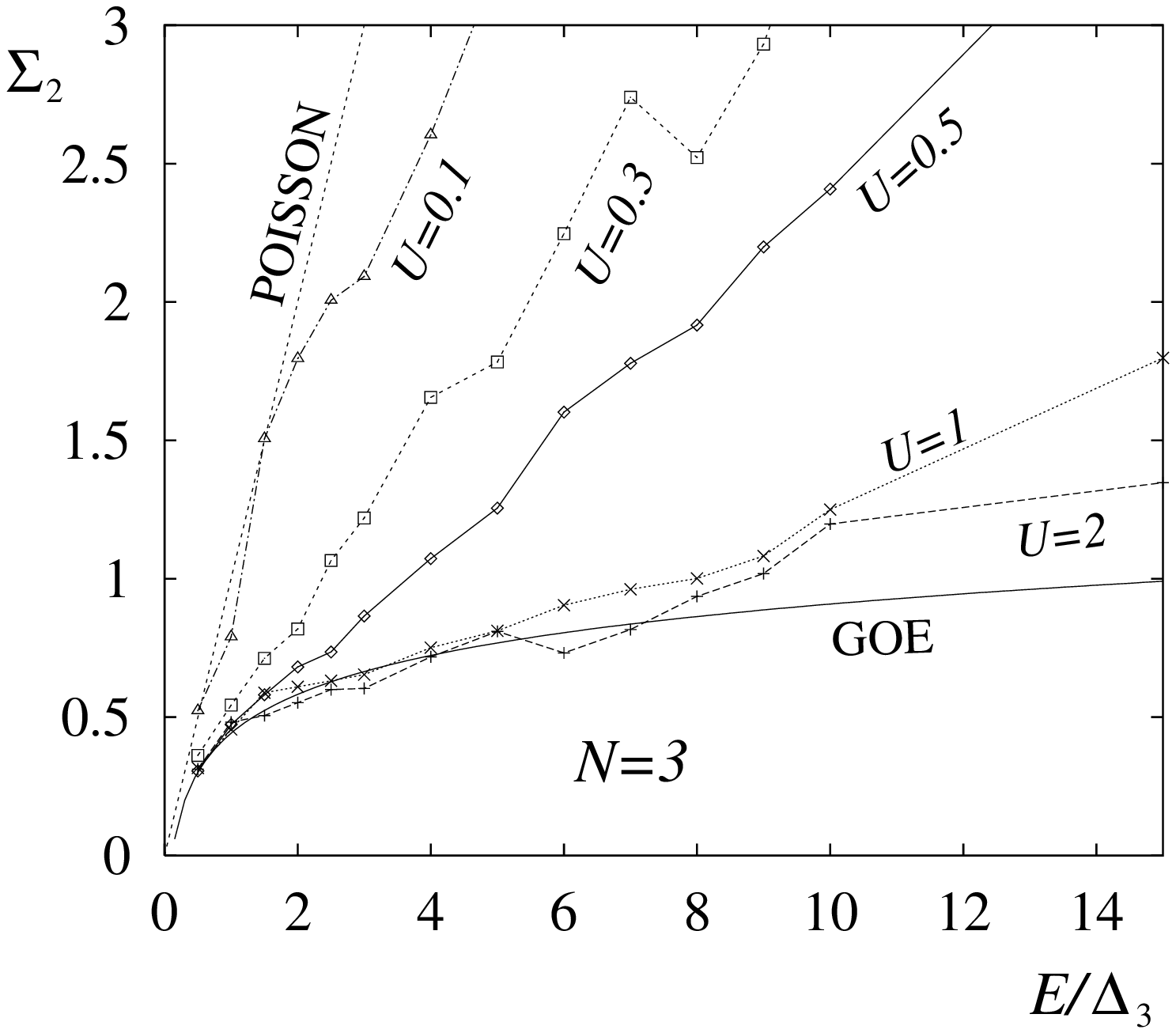}}
\caption[S2_23]{\label{S2_23}
$\Sigma_2 (E)$ for the spectrum of two (top) and three (bottom)
spinless Fermions in a $3\times 3 \times 3$ disordered lattice
($V=1,W=2$) for different values of $U$.}
\end{figure}
\noindent
more rigid around $U\approx 0.4$. This happens when neighboring states 
in energy (which usually are not directly coupled by the interaction) become 
correlated.

%============= P(s) =======================================

 This Poisson-Wigner transition can be studied more systematically by 
calculating the {\it local} fluctuation properties of the spectrum. The first 
quantity we look at is the level spacing distribution $P(s)$ in the center of 
the many-body spectrum around $E=0$. $P(s)$ for three particles is shown for a 
few values of $U$ in Fig.~\ref{PS3} (left). $\Delta$ is the corresponding mean 
level spacing ($\Delta_2$ or $\Delta_3$). At low $U$, $P(s)$ is quite close to 
the Poisson limit of uncorrelated levels, while it tends towards the universal 
Wigner result (GOE) at stronger $U$. This transition is described in a 
quantitative way by the integral
\begin{equation}
I:=\int_{1/2}^{2} ds \, P(s) \, ,
\end{equation}
which is shown in the right hand side picture of Fig.~\ref{PS3}. The transition
is much more abrupt for $N=3$ than for $N=2$. This agrees with the expected 
$U$-dependence of the Rabi frequency ($|U|$ for $N=2$ and $|U|^2$ for $N=3$) 
which characterizes the range of the level repulsion for weak $U$. 

%=============== effect on \Sigma_2 ========================

 The energy scale $E_{U}$ characterizing the universal spectral Wigner-Dyson 
rigidity can be extracted from the variance $\Sigma_{2}(E)=<N(E)^2> - <N(E)>^2$
of the number of energy levels in an interval of width $E$. Comparing the 
behavior of the spectrum to the GOE-behavior, one can identify this energy 
$E_{U}$ \cite{wp} (up to which the GOE-rigidity can be observed and above which
one can see 

\begin{figure}[tb]
\centerline{\epsfxsize=0.75\colwi\epsffile{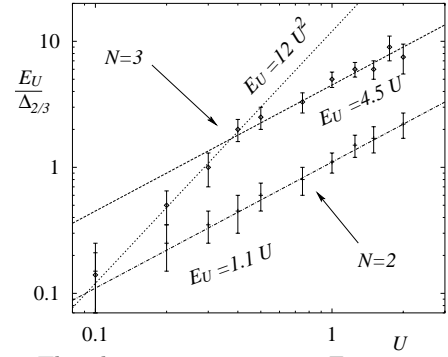}}
\caption[EU23]{\label{EU23}
The characteristic energy $E_{U}$ up to which the 2/3-particle spectrum
exhibits the universal Wigner-Dyson rigidity.}
\end{figure}
\noindent
significant deviations). $\Sigma_2(E)$ is shown in 
Fig.~\ref{S2_23}, near the band center. One can observe, for two as well as for
three particles, the Wigner-Dyson rigidity up to the energy scale $E_{U}$ where
the spectrum becomes less rigid.

%=============== EU ======================================

 For $N\ge 3$ particles, one can see that the characteristic energy $E_{U}$ 
does not coincide with the spread width $\Gamma_3$ of the eigenfunctions (as it
is the case for two particles when $\Gamma_2 > \Delta_{2}$ \cite{wp}). 

 For $N=3$, the crossover from the Poissonian behavior of uncorrelated levels 
to the universal behavior of the GOE is sharper. As can be seen in 
Fig.~\ref{EU23}, $E_{U}$ increases as $U^2$ for weak interaction. This 
corresponds to Rabi-oscillations due to second order coupling between nearby 
Fock states. This $|U^2|$-increase when $E_{U} < \Delta_3$ for $N=3$ is the 
analog of the $|U|$-increase observed for $N=2$ when $E_{U}<\Delta_2$. Above 
$U_{c1}$, $E_{U}$ seems to linearly increase as a function of $U$.

\subsection{Large $U$-limit}\label{numeric_Ular}

For large $U$ and small $M$, it is instructive to consider the Fock basis built
of the on-site orbitals. In this basis the 3-particle states, and the role of 
$U$, can be classified according to the number of next-neighbor configurations.
For a $3\times 3\times 3$ cubic lattice, we have 1746 out of the 2925 basis 
states where no next neighbor pairs occur and the Fock state is not shifted 
when the interaction increases. For 1008 basis states, there is one pair of 
particles nearby and the energy of the Fock states is $E(U=0)+U$. For the 
remaining 171 Fock states the particles are clustered such that their energy 
increases like $E(U=0)+2U$. Since the non-diagonal elements in this
representation are only due to one-particle kinetic energy which does not
depend on $U$, the different shifts of the Fock states lead to a splitting of 
the band into three parts, 1746 states around $E=0$, 1008 states around $E=U$ 
and 171 states around $E=2U$.

\begin{figure}[tb]
\centerline{\epsfxsize=0.95\colwi\epsffile{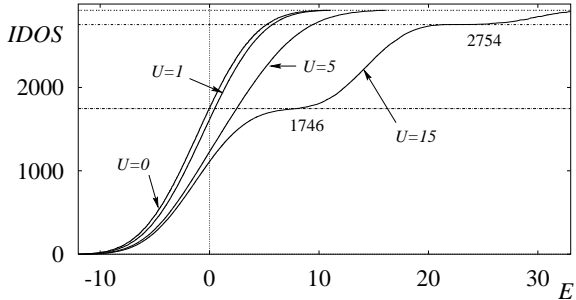}}
\caption[IDOS]{\label{IDOS}
The integrated density of the three-particle states for different values of
the interaction.}
\end{figure}

This can be seen for $U=15$ in Fig.~\ref{IDOS}, where the integrated density of
states $IDOS=\int_{-\infty}^{E}dE'\rho(E')$ is plotted for one realization of 
the disorder and different

\noindent
values of the interactions. The density of states can be fitted 
by a sum of three Gaussians centered at $E=0$, $U$, and $2U$ with
weights corresponding to the above mentioned numbers of occurrence of these 
shifts when $U\approx 15$. For $U\approx 1$, the interaction induced gaps in 
the spectrum are totally removed by the one-particle kinetic contributions.

\subsection{Evidence for the existence of two thresholds}

The numerical results for the three-particle spectrum clearly exhibit two 
different characteristic interaction strengths. First, the local fluctuations 
of the level spacing changes from Poisson to GOE when $U\approx 0.3$, as it can
be seen from Fig.~\ref{PS3}. This is consistent with the results obtained for 
$\Sigma_2$ where $E_{U} \approx \Delta_3$ when $U\approx 0.3$. Thus, adjacent 
energy levels are correlated when $U>0.3$. This gives $U_{c1}\approx 0.3$.
Below $U_{c1}$, $E_{U}$ increases as $U^2$ in agreement with Rabi-oscillations
in the perturbative regime. Above $U=U_{c1}$, $g_3 = E_{U}/\Delta_3 >1$ and the
perturbation theory breaks down. An indication for this is that $E_{U}\propto 
U$ and not $E_{U} \propto \Gamma_3^{\rm (i)} \propto U^4$, as implied by the 
perturbation theory. Though we have no explanation for this linear behavior, we
emphasize that this absence of a $U^4$-behavior gives a strong hint that the 
range of validity of perturbation theory is limited to 
$U_{\rm typ} \leq \Delta_2^{\rm eff}$.

Note that nothing striking is observed in the behavior of the total spread
width $\Gamma_3$ of the states at $U=U_{c1}$. There seems to be a crossover 
from $R\propto U$ to $R\propto U^2$ in the participation ratio $R$, but since
$R=1$ at $U=0$, it might also be the signature of a saturation which is not 
connected to a characteristic energy scale but which becomes irrelevant when 
$R\gg 1$.

However, there is something interesting happening at $U\approx 1.5$ where both,
$\Gamma_3$ and $R$ undergo a transition from a regime where they increase as 
$U^2$ to a regime where they increase as $U^{3/2}$. This is clear evidence for 
a change in the structure of the eigenstates as it is expected from our 
theoretical considerations at $U=U_{c2}$. At this strength of the interaction,
we expect the indirect spread width $\Gamma_3^{\rm (i)}$ to be of the order of 
$\Delta_2^{\rm eff}$. The problem is now to estimate $\Gamma_3^{\rm i}$ when 
$U>U_{c1}$ since the perturbation theory does not work. If one uses the 
numerical result for $ E_{U} \approx \Gamma_3^{\rm i}$ in this non-ergodic 
Wigner-Dyson regime, one gets for three particles $U_{c2}-U_{c1} \approx 0.8$ 
since $\Delta_2^{\rm eff}\approx 3.3 \Delta_3$.

 In summary, we find a clear evidence of the existence of two distinct 
thresholds $U_{c1}$ and $U_{c2}$, as well as a strong indication that the 
self-energy of an individual Fock state cannot be evaluated by perturbation 
theory for $g_N \geq 1$.

\section{Implication for quantum localization in real space}

 Very recently, scaling-type concepts have been applied to two particles with a
local interaction. When Shepelyansky \cite{shepelyansky} had pointed out that 
in insulating systems certain two-particle wave-functions could be delocalized 
with respect to the one-particle states, Imry extended the Thouless 
Block-scaling picture \cite{imry} and introduced a ``two-particle conductance''
$g_2 = \Gamma_{\rm U}/\Delta_2$, where $\Gamma_{\rm U}$ is the decay rate of 
the states in boxes of size $L_1$ due to the interactive coupling to other 
boxes. This $\Gamma_{\rm U}$ is identical to the $\Gamma_2$ characterizing the 
spectral fluctuations for a given block. $\Delta_2$ is the mean spacing of the 
two-particle spectrum in a block. Using Fermi's golden rule for the estimate of
$\Gamma_{\rm U}$, this yields a pair localization length $L_2 \propto U^2 
L_1^2$, in agreement with the results of Shepelyansky. The existence of this 
delocalization effect has been confirmed in numerical studies 
\cite{fmgpw,wmgpf,oppen}. 

 While in this paper we have considered the metallic phase, away from the 
localization transition, our treatment may form the basis for finding the 
localization length for an $N$-particle system (where $N \ge 3$). This 
localization length too is increased for large enough interactions. The scaling
theory prescription for obtaining this localization length is straightforward, 
in principle. One has to increase  the system size, $L$, and watch the 
$L$-dependence of both the {\it effective} spacing $\Delta_N^{\rm eff} $ of the
$N$-particle levels which are re-organized by the interaction when $L> L_1$ and
the energy $E_U$ above which the Wigner-Dyson rigidity does not apply for this 
subset of levels. The $L$ at which $E_U / \Delta_N^{\rm eff}$ becomes of order
unity is the N-particle localization length. The parametric dependence embodied
in our estimate of $g_3$ ($g_N$) can be used for this purpose. We expect the 
delocalization to become even stronger for $N \ge 3$. It is tempting to 
suggest that such effects are at the origin of the recent 
observations~\cite{kravchenko,popovic} of a two dimensional metallic phase 
driven by the interactions in Si-MOSFET.  

\section{Quasi-particle lifetime and localization transition in the
Fock-space}

 We conclude by stressing the analogies and the differences between this study 
and the recently proposed~\cite{agkl} approach to quasi-particle lifetime in an
isolated system. In these two studies, two characteristic energies are 
identified and a transition appears when their ratio is of order unity. This is
the transition from weak perturbative mixing to the golden-rule decay, as 
mentioned in the introduction. This is also the threshold where the 
perturbation theory in $U$ or $\epsilon$ breaks down. In this sense, the two 
studies use very similar concepts, but the considered characteristic scales are
not the same. This is because two different situations were considered 
(quasi-particles in a finite-density Fermi gas, vs. a dilute gas of "real" 
particles). It may be argued, as in Ref.\cite{imry}, that the difference 
between these two systems  is mainly in the counting of the densities of 
excited states, and that in principle both could be treated by similar scaling
considerations.

We have mainly discussed what is the ratio (Thouless number) which controls the
transition from Poisson to Wigner in the bulk of the $N$-body spectrum. The 
threshold $U_{c_1}$ corresponds to $U_{\rm typ} \approx \Delta_2^{\rm eff}$. 
Therefore, for very few particles where $\Delta_2^{\rm eff} \approx \Delta_2$, 
and $g_n\approx g_2\approx 1$ at $U_{c1}$, this first threshold is eventually 
related to the ratio $g_2$ of the two-particle decay width $\Gamma_2$ over the 
two particle spacing $\Delta_2$. This is not due to the fact that $\Delta_N$ is
not a relevant energy scale, but because this is the contribution of order 
$P=N/2 $ or $(N+1)/2$ (depending on the parity of $N$) in the perturbation
theory of the decay width $\Gamma_N$ which matters for the spectral
fluctuations at the scale $\Delta_N$. Therefore, as previously proposed by 
Shepelyansky and Sushkov, we just need to have the two particle levels well 
coupled by the interaction in order to have the same thing for the $N$-body 
levels. In this case the conditions for the establishment  of Wigner-Dyson 
rigidity and to have an effective decay are similar and both are 
$U_{\rm typ} \sim \Delta_2^{\rm eff}$.

 For the lifetime of quasi-particles in a zero-dimensional Fermi system, the 
relevant  ratio is made from the two different scales: the decay width of a 
single quasi-particle $\Gamma_{\rm sp} (\epsilon)$ (resulting from the 
disintegration of a single quasi-particle into two quasi-electrons and a hole) 
and the accessible three quasi-particle level spacing $\Delta_3(\epsilon)$. In 
this connection, a quasi-particle is considered at an excitation energy
$\epsilon$ above the Fermi energy of an isolated system. The corresponding
Fermi vacuum is assumed to always provide a new electron-hole pair at each 
interaction process, such that the relevant decay width is not the one 
characterizing the decay of a certain Slater determinant to those with the same
quasi-particle content, but to those with a quasi-particle content increased by
a new electron-hole pair. It should be emphasized that this does not correspond
to the dilute limit discussed in our work, but to a limit of finite density of
particles, such that the Fermi vacuum can be considered as an unlimited 
reservoir of particle-hole excitations. $\Gamma_{\rm sp} (\epsilon) 
\approx U_{\rm typ}^2 / \Delta_3(\epsilon) \approx \Delta_3(\epsilon)$ defines 
the (second) excitation threshold $\epsilon^{*}$ in Ref.\cite{agkl}. The 
threshold is obtained when the typical magnitude of the interaction matrix 
element $U_{\rm typ}$ is of order $\Delta_3(\epsilon)$, unlike the value 
$\Delta_2^{\rm eff}$ which was relevant  in our dilute limit.  In 
Ref.\cite{agkl}, it is suggested that the two problems of level statistics and 
golden rule decay are unrelated: that delocalization in Fock space does not 
mean that the spectrum should have Wigner-Dyson statistics. 
For a Cayley tree, this disagrees with the results of recent supersymmetric 
calculations\cite{mirlin_fyodorov} using a non-linear sigma model 
formulation. For the "dilute" case of a small number 
of particles in a large volume, it follows from our work that these two 
properties are very intimately related. 
\par
 In this work, the transition\footnote{This study is clearly restricted to 
disordered systems, where the one body states are chaotic. In clean systems, 
the structure of the interaction in the Fock basis built out from plane waves 
is very different (see for instance Ref.\cite{ponomarev,waintal}). However, 
disorder is not necessary for having chaotic $N$-body states and Wigner-Dyson 
rigidity. When the one body problem is integrable, the condition suggested by 
many numerical studies~\cite{montambaux2} for having chaos is the 
nonintegrability of the strongly correlated fermion system.} from an 
uncorrelated spectrum to a fully ergodic one seems to occur in two stages: 
$U_{c1}$ and $U_{c2}$. We do not have a qualitative understanding of the 
behavior in the intermediate regime. Whether this regime will survive with 
increasing number of particles, and what is the precise r
elationship\footnote{After completion of this manuscript, we have noticed 
related works~\cite{js,ab} where the relevance of our first threshold $U_{c1}$ 
for the onset of chaos is studied, particularly in the case of low energy 
quasi-particles excited above the Fermi sea.} with Ref.\cite{agkl} are 
questions which deserve further investigation. 
  
%================ Acknowledgment ==========================
We are grateful to Yuval Gefen, Alex Kamenev and Dima Shepelyansky
for instructive discussions.
Dietmar Weinmann  acknowledges the financial support of the
European HCM program. Research at the Weizmann Institute was supported
by the Fund for Basic Research administered by the Israel Academy of Sciences
and by the German-Israeli Foundation (GIF), Jerusalem.

%          ================= REFERENCES =================
%

\end{document}